**Physics-informed acquisition weighting for stoichiometry-constrained Bayesian optimization of oxide thin-film growth**

Yuki K. Wakabayashi,[1,*] Takuma Otsuka,[2] Yoshiharu Krockenberger,[1] and Yoshitaka Taniyasu[1]

[1]*Basic Research Laboratories, NTT, Inc., Atsugi, Kanagawa 243-0198, Japan*
[2]*Communication Science Laboratories, NTT, Inc., Soraku-gun, Kyoto 619-0237, Japan*

[*]Corresponding author: yuuki.wakabayashi@ntt.com

**Abstract**
We present a physics-informed Bayesian optimization (PIBO) with a concise modification to its acquisition function to incorporate the physical prior knowledge. Specifically, this method multiplies the expected improvement (EI) by a weight encoding prior crystal growth physics. When applied to $LaAlO_3$ molecular-beam epitaxy, the weighting function defines a flat stoichiometric window and penalizes off-window proposals, thereby steering the optimization toward physically plausible regions while maintaining controlled exploration. In a closed-loop optimization, relative to the bare EI, which often proposes off-stoichiometric conditions, the weighted EI constrains the search toward stoichiometric regions while retaining sufficient flexibility to explore neighboring conditions, eventually identifying an optimum slightly beyond the stoichiometric window. Within only 15 growth runs, the lattice constant of the grown $LaAlO_3$ film converged to the bulk value, evidencing efficient and rapid optimization for the ideal stoichiometric growth. Because physics knowledge is incorporated solely through the weighting function, the approach requires only minimal modification to standard BO workflows and is readily applicable to other material systems, offering a general and practical route to AI-driven materials synthesis.

**Introduction**

Recent advances in artificial intelligence (AI) and machine learning (ML) have enabled data-driven approaches to accelerate materials discovery and process optimization.[1,2,3,4,5,6,7,8,9,10,11] In thin-film synthesis, Bayesian optimization (BO) has efficiently explored multidimensional growth-parameter space and uncovered hidden correlations between the growth conditions and material properties,[12,13,14,15] an approach often termed ML-assisted thin-film growth.[16,17] Building on this foundation, physics-informed Bayesian optimization (PIBO), which incorporates crystal-growth physics into the algorithm, has recently emerged as a promising approach for improving the efficiency and accuracy of compound-semiconductor epitaxy.[18] Recent PIBO advances often embed physics knowledge in the surrogate model to improve the prediction of physical characteristics, for example, through physics-informed kernel design[19] or by enhancing extrapolation via residual learning atop a physics-based model.[18] Alternatively, a modification to the acquisition function can provide an intuitive and interpretable way to incorporate physics into BO with minimal changes to existing pipelines, thereby facilitating straightforward domain transfer. The development of optimization frameworks integrating ML models and physics knowledge[19,20,21,22] is becoming increasingly important across diverse areas of materials science, from semiconductor processing to complex oxide epitaxy.

One of the most critical challenges in the epitaxial growth of complex oxides is achieving precise stoichiometric control.[23,24,25,26,27] Even slight deviations from ideal cation ratios will result in point defects, dislocations, secondary phases, and structural distortions, which strongly influence the electronic and optical properties. In oxide thin-film growth, such as molecular-beam epitaxy (MBE), stoichiometry control typically relies on empirical tuning of growth parameters such as flux ratios and substrate temperatures. These parameters are often coupled in complex and nontrivial ways, making manual optimization time-consuming, operator-dependent, and unreliable. Therefore, incorporating compositional constraints into data-driven optimization frameworks is essential for guiding experiments toward physically reasonable and defect-free growth regimes.

Here, we present a PIBO framework that multiplies the acquisition function by a weight that encodes prior crystal growth physics. We demonstrate this approach in $LaAlO_3$ (LAO) MBE under stoichiometry constraints, a representative complex oxide where precise cation control is critical.[26] The weighting encodes thermodynamic priors on phase stability, here represented by the nominal La:Al supply ratio of 1:1, thereby steering the optimization toward chemically balanced compositions. This approach requires only a minor modification to standard BO implementations, making it straightforward to implement in existing experimental workflows. Using this weighted EI scheme, the lattice constant estimated by X-ray diffraction converged to the bulk value after only 15 iterative MBE growth cycles, accompanied by significant improvements in crystalline quality. These results highlight the effectiveness of the proposed PIBO framework in accelerating thin-film growth optimization and enhancing film quality in complex oxides, while also being readily extensible to other materials and autonomous synthesis systems.

**Results**
**PIBO with stoichiometry-constrained weighting**
We implemented a PIBO framework that incorporates a stoichiometry-based constraint directly into the acquisition. Figure 1 summarizes the closed-loop workflow of the proposed PIBO, indicating where physics knowledge is introduced into the acquisition via the weighting function. BO is a method for optimizing a black-box function of the form $y = f(x)$,[28] where the function $f$ is unknown and costly to evaluate, given a $D$-dimensional input $x \in \chi \subset R^D$ within a defined search space. In materials growth optimization, $x$ and $y$ represent the growth parameters and physical properties used to evaluate grown materials, respectively. Examples of physical properties include electrical resistance and X-ray diffraction intensity. A Gaussian-process (GP) regression with a Matérn kernel constructs a prediction of $f(x)$ as a Gaussian-distributed variable based on the past $n$ observations, $D_n = \{(x_i, y_i)\}_{i=1}^n$, and the acquisition function of weighted EI is maximized at each iteration to select the next experiment. We start from several random initial points and proceed sequentially until a fixed evaluation criterion is reached. Except for the stoichiometry-constrained EI weighting introduced in this work, all remaining components follow our previous implementation, including the adaptive (random) prior mean to escape local minima[29] and the imputation of failed experiments with the lowest observed evaluation so as not to artificially shrink the search space.[15] Detailed equations and hyperparameter settings are provided in Ref. [29].

We adjust the EI by applying a multiplicative weight that encourages sampling near the stoichiometric center $c$ of the La/Al ratio ($c = 1$ in this study). The choice of $c = 1$ is physically motivated by the fact that oxide species such as $La_2O_3$ and $Al_2O_3$, which are likely to form during LAO film growth, have low vapor pressures.[30,31] The low vapor pressures indicate that the sticking coefficients of La and Al are close to unity under our growth conditions. It is important to design the weighting function appropriately for each material system, based on its specific physics knowledge of crystal growth. For example, for material systems such as $SrRuO_3$, the formation of volatile oxide species like $RuO_2$ is expected during growth, and desorption or re-evaporation processes become pronounced.[15,32,33] In such a case, assigning higher weights to Ru-rich regions can compensate for the preferential loss of volatile components.

Let $\delta$ denote the nominal La/Al supply ratio at a candidate point $x$. The weight $w(\delta)$ is defined by a flat window of half-width $\tau$ around $c$ and a Gaussian decay outside the window:

$$w(\delta) = \begin{cases} 1, & |\delta - c| \leq \tau, \\ \exp\left(-\frac{(|\delta - c| - \tau)^2}{2\sigma^2}\right), & |\delta - c| > \tau. \end{cases} \quad (1)$$

The stoichiometry-constrained acquisition (= weighted EI) is then

$$\alpha_{\text{SC}}(\delta) = [\epsilon + (1 - \epsilon)w(\delta)]\text{EI}(\delta). \quad (2)$$

The parameter $\tau$ defines a flat window $|\delta - c| \leq \tau$ where EI remains unmodified. A smaller $\tau$ concentrates exploration tightly around the stoichiometric center, but makes the search more susceptible to drifting away from the true optimum due to experimental noise and inaccuracies in the physics model. The decay width $\sigma$ controls how fast the weight falls outside the window; larger $\sigma$ maintains exploration in off-stoichiometric regions, whereas smaller $\sigma$ penalizes them more sharply. The parameter $\epsilon \in [0,1]$ is a floor factor that preserves a nonzero probability of exploring off-stoichiometric regions, which improves robustness to model misspecification but slows concentration around $c$. In practice, the parameter choices must reflect the experimental accuracy of the nominal elemental supply ratios and the reliability of the physics model describing the crystal-growth process. In this study, we set the parameters to $\tau = 0.10$, $\sigma = 0.10$, and $\varepsilon = 0.30$ for all optimization runs [Fig. 2(a)]. For reference, Fig. 2(b) shows an example of a relaxed stoichiometry constraint realized by increasing σ to 0.4. The reduced rate of decay for the weighting function outside the flat window permits broader exploration of off-stoichiometric regions, yet still maintains preferential sampling near the stoichiometric center.

    The present formulation in Eqs. (1) and (2) can also be positioned within the broader context of Bayesian optimization strategies. Our method employs the weight function to modulate the acquisition function to reflect the stoichiometry constraint. Weight functions have been applied to EI in constrained optimization setups,[34,35,36] where they are typically designed as the probability that constraints are satisfied. This probability is predicted by a separate model as a function of the $D$-dimensional input variables and is iteratively updated as more observation data are collected. In contrast, our weighting function is not data-driven but rather designed to emphasize a physically meaningful region of the search space, directly encoding prior knowledge of the growth process. We also note that the use of $\varepsilon$ in our acquisition function of Eq. (2) is similar to the $\varepsilon$-greedy strategy in the literature on the multi-armed bandit problem and reinforcement learning.[37] The $\varepsilon$-greedy strategy makes a random decision with probability $\varepsilon$ to encourage exploration. Our approach follows the idea of promoting explorative behavior by smoothing the weight function with parameter $\varepsilon$. Our method recovers the original EI-based optimization by increasing $\varepsilon$ to 1, rather than resorting to a random search process.

    Here, we describe the flexible extension of the formulation in Eq. (1). For materials where the volatility asymmetry is significant, such as $SrRuO_3$, an asymmetric weighting can be defined to favor one side of the stoichiometric center. In this case, $\delta$ is defined as the Sr/Ru supply ratio. For instance, when Ru-rich conditions are desirable, the weight may be modified as

$$\begin{cases} 1, & |\delta - c| \leq \tau, \\ \exp\left(-\frac{(|\delta-c|-\tau)^2}{2\sigma_1^2}\right), & \delta - c > \tau. \\ \exp\left(-\frac{(|\delta-c|-\tau)^2}{2\sigma_2^2}\right), & \delta - c < -\tau. \end{cases} \qquad (3)$$

where $\sigma_1$ and $\sigma_2$ control the decay widths on the Ru-poor and Ru-rich sides, respectively. By setting a larger decay width on the Ru-poor side than on the Ru-rich side ($\sigma_1 < \sigma_2$), the exploration can be preferentially directed toward the Ru-rich side. Alternatively, by taking $\sigma_2 \to \infty$, the Ru-rich side can be made completely unpenalized. As another example, to favor Ruddlesden–Popper series of $Sr_{n+1}Ru_nO_{3n+1}$,[38] stoichiometric center $c$ should be $c_n = (n + 1)/n$. In these ways, this general formulation allows the acquisition to be biased according to the expected volatility behavior and the target stoichiometry of specific elements. Furthermore, as a future extension, introducing weighting functions for temperature and oxygen partial pressure according to thermodynamic stability would be a promising way to incorporate thermodynamics, such as thermodynamic data, free energy of formation of related compounds, and Ellingham diagrams,[15,32,33] into the framework.

**Growth optimization by PIBO**

We applied the PIBO method with stoichiometry-constrained weighting to the MBE growth of LAO films to demonstrate its practical applicability and effectiveness (see Methods section "MBE growth of LAO films, XRD, and AFM measurements" for details of experimental setups). LAO is a prototypical wide-band-gap perovskite oxide[39,40,41] widely used as a microwave dielectric and as a platform for thermoelectric/catalytic heterostructures and oxide electronics. Maintaining stoichiometric LAO is of pivotal industrial importance; previous studies show that oxygen vacancies cause sub-gap absorption,[42] that a slight La/Al imbalance suppresses the electron gas formation at LAO/SrTiO$_3$ (STO) interfaces,[25] and that such off-stoichiometries shift the pseudocubic lattice constant away from the bulk value[26] and degrade dielectric properties,[43] highlighting the need for growth strategies that stabilize stoichiometry within the narrow layer-by-layer growth window.[27]

To evaluate the stoichiometry of the LAO films, we measured $\theta$-$2\theta$ X-ray diffraction (XRD) of the LAO films grown on STO (001) substrates since an increase in the lattice constant serves as a reliable indicator of off-stoichiometry in LAO, arising from variations in cation and/or oxygen concentration.[26,27,44] Therefore, we used $\Delta c$ as the evaluation value, defined as the difference between the $c$-axis lattice constant of the film, determined by the Nelson–Riley extrapolation method, and that of bulk LAO. Note that LAO films grown on STO (001) substrates are known to be fully relaxed when their thickness exceeds approximately 4 nm due to the large lattice mismatch between LAO (3.821 Å) and STO (3.905 Å).[45] As a result, the $c$-axis lattice constant directly reflects the pseudocubic lattice constant of the LAO films, allowing a straightforward comparison with bulk values. When XRD peaks from the LAO phase were indiscernible, the evaluation value for those samples was assigned to be the worst experimental $\Delta c$ value at that time. By imputing the missing data generated upon the non-formation of the designated phase, the system achieved continuous exploration throughout the extensive three-dimensional parameter space.[15]

We initialized the search using five randomly selected growth conditions as preliminary data, and then ran the PIBO sequence. At each iteration, we compute $\alpha_{SC}(x)$ (weighted EI) with $x = (\delta$, growth temperature, ozone-nozzle-to-substrate distance) and the next experiment was chosen by maximizing the $\alpha_{SC}(x)$. The weight depends solely on $\delta$, while the GP surrogate and weighted EI are defined over the three-dimensional space. Figure 3 illustrates how the choice of acquisition function (weighted EI and bare EI) affects the optimization process, with examples of the GP prediction and the corresponding next-experiment proposals after nine observations. The GP model prediction of $\Delta c$ [Fig. 3(a)] displays a low-value valley around the stoichiometric center ($\delta = 1$), in agreement with prior knowledge of LAO crystal growth. In weighted EI [Fig. 3(b)], exploration is preferentially focused near the stoichiometric flat window while still leaving room along the ozone-nozzle-substrate distance axis, indicating remaining in-plane freedom for exploration. By contrast, the bare EI [Fig. 3(c)] attains high values outside the stoichiometric flat window, which would lead to experimental failure or high $\Delta c$ values.

After nine observations, the lowest experimental $\Delta c$ at that time was 0.16 Å. The next-experiment conditions suggested by the weighted EI [($\delta$, growth temperature, ozone-nozzle-to-substrate distance) = (0.88, 892°C, 5.5 mm)] yielded a film with $\Delta c$ = 0.03 Å, representing a significant improvement, whereas the suggested conditions from the bare EI (0.75, 892°C, 5.5 mm) resulted in disappearance of the LAO diffraction peaks, indicating failure of phase formation. For comparison, the bare-EI prediction was evaluated at the same temperature (892 °C) where the highest weighted-EI value was obtained. We also performed additional experiments using a more relaxed weighted-EI function shown in Fig. 1(b), in which the stoichiometric constraint was incorporated more gently. In this case, the relaxed weighted EI proposed conditions (0.84, 892°C, 6.0 mm) [Fig. 3(d)] between those suggested by the weighted and bare EIs, and yielded a film with $\Delta c$ = 0.08 Å, showing an improvement relative to the bare-EI result while maintaining moderate exploration.

Figure 4(a) plots the $\Delta c$ values across the three-dimensional parameter space for 15 observations. As explicitly shown in the corresponding projection onto the $\delta$-temperature plane [Fig. 4(b)], the acquisition weighting prioritizes near-stoichiometric conditions while retaining a nonzero exploration probability elsewhere. The observations progressively concentrate near the stoichiometric flat window of $0.9 \leq \delta \leq 1.1$, and conditions far from the flat window tend to yield larger $\Delta c$ or failures. Figure 4(c) shows $\Delta c$ as a function of the growth run number. Despite occasional failures or temporary degradations in $\Delta c$, the lowest $\Delta c$ decreases with increasing growth number and reaches $\Delta c = 0$ Å after 15 runs, demonstrating rapid convergence of the lattice constant to the bulk value. The ideal $\Delta c = 0.00$ Å was achieved at $\delta = 0.88$, growth temperature = 892°C, and ozone-nozzle-to-substrate distance = 5.5 mm. While preserving a nonzero sampling probability outside the flat stoichiometric window, the optimization eventually identified an optimum just beyond it, plausibly reflecting small calibration offsets or non-unity sticking coefficients. Notably, the ideal growth with $\delta = 0.88$ means that our method is advantageous over a naive approach that simply restricts the search range to $0.9 \leq \delta \leq 1.1$, equivalently setting $\varepsilon = 0$ and $\sigma \to 0$ in our method, in order to implement the stoichiometry constraint. These results indicate a balanced trade-off between exploration and exploitation, demonstrating that the stoichiometry-constrained acquisition efficiently concentrates experiments near the desired composition while remaining robust to noise and modest inaccuracies in the prior growth physics.

**Crystallographic properties of stoichiometric LAO films**

We experimentally characterized the stoichiometric LAO films ($\Delta c = 0$ Å). Figure 5(a) presents the XRD $\theta$-$2\theta$ scans of the optimized ($\Delta c = 0$ Å) and unoptimized ($\Delta c = 0.032$ Å) LAO films grown on STO (001) substrates. The optimized film not only exhibits a lattice constant identical to that of bulk LAO, but also shows more than a tenfold increase in the LAO (002) peak intensity relative to unoptimized films, clearly indicating markedly improved crystallinity. The XRD $\theta$-$2\theta$ scan of the optimized LAO film homoepitaxially grown on an LAO (001) substrate is shown in Fig. 5(b). No additional peaks corresponding to precipitates or secondary phases are observed, confirming the single-crystalline nature of the homoepitaxial film. Apart from weak Laue thickness fringes, the film peaks exactly coincide with the substrate LAO peaks and thus are indistinguishable, fully consistent with $\Delta c = 0$ Å. Laue fringes arise from interference between X-rays reflected at the film surface and at the interface; their presence typically signifies uniform thickness and an atomically smooth film surface and interface.

Epitaxial growth of the high-quality single-crystalline homoepitaxial LAO film is also verified by the high angle annular dark-field (HAADF) and annular bright-field (ABF) scanning transmission microscopy (STEM) (Fig. 6). The atomically smooth surface of the optimized LAO film on an STO (001) substrate was further confirmed by atomic force microscopy (AFM) (Fig. 7). The AFM image of a 2-nm-thick LAO film grown on an STO(001) substrate shows a step-and-terrace structure with a step height of approximately 0.4 nm (corresponding to a single unit-cell height) and a low root-mean-square roughness of 0.16 nm, indicating two-dimensional layer-by-layer growth. Notably, the 2-nm thickness is below the reported critical thickness for misfit relaxation in LAO/STO, and the film is thus expected to remain coherently strained.[45]

**Discussion**

The proposed PIBO framework offers broad versatility as a platform for integrating physics knowledge of material synthesis into statistical ML. Its flexible weighting formulation allows researchers to encode prior knowledge, such as volatility, stoichiometric balance, or thermodynamic stability, directly into the acquisition process without altering the underlying optimization logic. This design efficiently biases exploration toward physically reasonable regimes while preserving a finite probability of sampling outside the preferred window, thereby maintaining robustness against imperfect prior physics models. This balance between constraint and flexibility provides a promising approach that leverages prior knowledge without excluding the discovery of complex growth behaviors and unexpected synthesis pathways. It is particularly compatible with autonomous synthesis systems that continuously explore and learn through iterative experimentation.

In summary, we presented a PIBO framework with stoichiometry-constrained weighting for accelerating the MBE growth optimization of LAO thin films. The acquisition function was modified by a multiplicative weight tied to the stoichiometric La/Al ratio, based on prior knowledge of crystal growth physics. This weighting ensures that proposals are preferentially drawn from the stoichiometric flat window while retaining controlled exploration of other parameters. The optimization converged efficiently in practice: after only 15 growth runs, the lattice constant of the LAO films matched the bulk value, evidencing stoichiometric growth. The optimized films exhibited clear hallmarks of improved crystalline quality, including more than a tenfold increase in the LAO XRD peak intensity compared with the unoptimized film. These results demonstrate that the proposed PIBO scheme efficiently concentrates experiments in physically reasonable regions and improves structural quality, offering a practical route toward AI-assisted materials synthesis.

**METHODS**
**MBE growth of LAO films, XRD, and AFM measurements**

Epitaxial LAO films with a thickness of 50 nm were grown on STO (001) and LAO (001) substrates in a custom-designed molecular beam epitaxy (MBE) system equipped with multiple e-beam evaporators. Detailed information about the MBE system is described elsewhere.[46] The La and Al elemental fluxes were controlled by monitoring the flux rates with an electron-impact emission spectroscopy sensor and feeding back the measured values to the power supplies of the e-beam evaporators.[16,47] The oxidation during growth was carried out with ozone ($O_3$) gas (~15% $O_3$ + 85% $O_2$) introduced through an alumina nozzle directed toward the substrate. For stoichiometric LAO growth, fine-tuning of the growth conditions (the ratio of the La flux to the Al flux $\delta$, growth temperature, and local ozone pressure at the growth surface) was critical. To vary the La-to-Al flux ratio, we changed the La flux while keeping the Al flux constant at 0.3 Å/s. The growth temperature was controlled using an Ir heater. The local ozone pressure at the growth surface was adjusted by changing the ozone nozzle–substrate distance while maintaining the $O_3$ gas flow rate at approximately 2 sccm. The search ranges for the $\delta$, growth temperature, and ozone-nozzle-to-substrate distance were 0.5-1.5, 600-900°C, and 5-25 mm, respectively.

XRD measurements were performed using a Bruker D8 Discover. The X-ray wavelength was 1.54060 Å (Cu K$\alpha_1$ line source) with a germanium (220) monochromator. AFM measurements were performed using a Bruker Dimension FastScan AFM. STEM measurements were performed using a JEOL JEM-ARM 200F microscope.

**AUTHOR CONTRIBUTIONS**
Y.K.W. conceived the idea, designed the experiments, and directed and supervised the project. Y.K.W. and T.O. designed and implemented the physics-informed Bayesian optimization algorithm. Y.K.W. and Y.K. carried out the MBE growth and characterization of the samples. Y.K.W. wrote the paper with input from all authors.

**COMPETING INTERESTS**
The authors declare no competing interests.

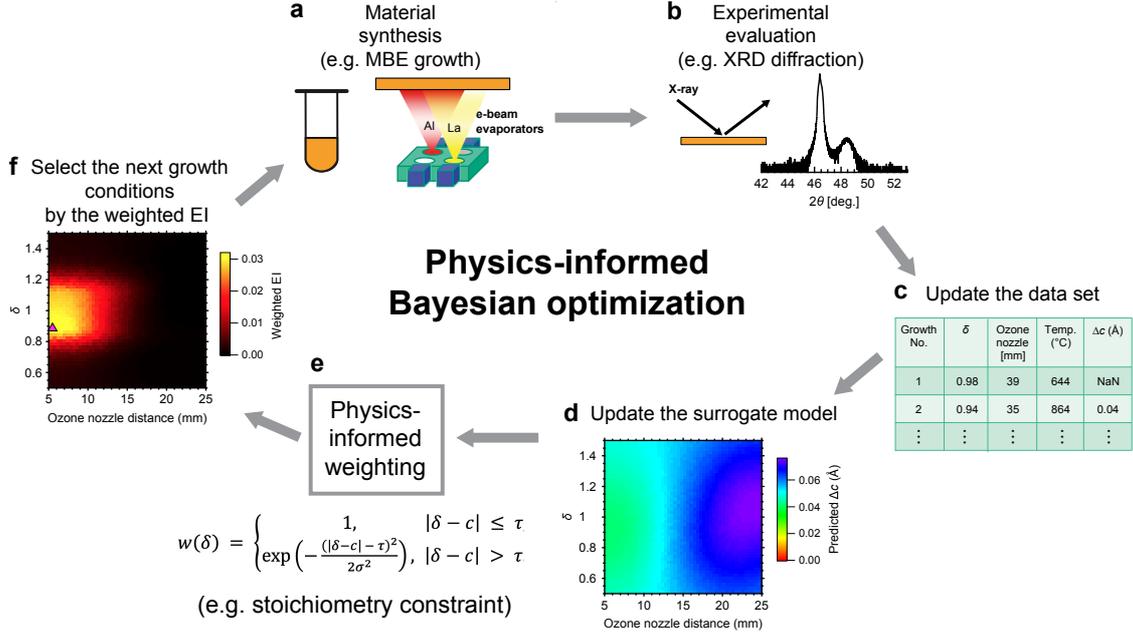

**Fig. 1. Schematic workflow of the PIBO. a** Synthesis: grow materials under controllable parameters. **b** Evaluation: obtain the evaluation value of the grown material. **c** Data set update: append the newest growth parameters and evaluation to the data set. **d** Surrogate prediction: update the surrogate model using the data set. **e** Physics-informed weighting: apply a weighting based on prior physics knowledge to a standard EI acquisition function. **f** Next-experimental proposal: Select the next growth conditions where weighted EI is maximized.

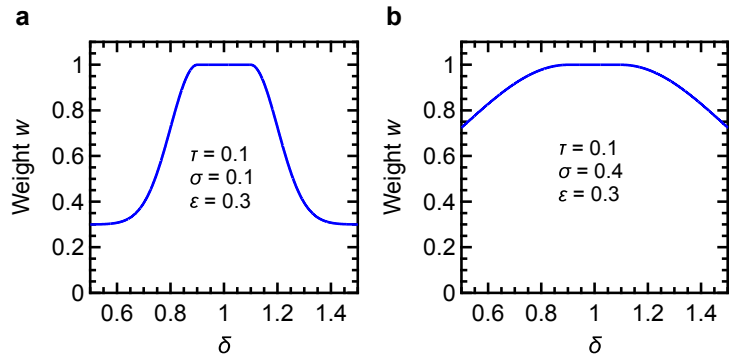

**Fig. 2. Weighting functions.** $[\varepsilon + (1 - \varepsilon)\, w(\delta)]$ as a function of the La-to-Al flux ratio δ, showing the unity plateau for $|\delta - c| \leq \tau$ and the Gaussian roll-off for $|\delta - c| > \tau$ with the parameters of **a** $\tau = 0.1$, $\sigma = 0.1$, and $\varepsilon = 0.3$, and **b** $\tau = 0.1$, $\sigma = 0.4$, and $\varepsilon = 0.3$.

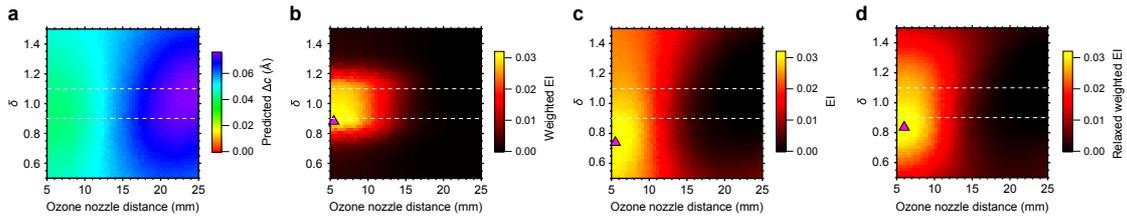

**Fig. 3. Prediction results.** Two-dimensional plots of **a** predicted $\Delta c$ values, **b** weighted EI, **c** bare EI, and **d** relaxed weighted EI for nine observations at the temperature of 892°C, at which the highest weighted EI value was obtained. The white dashed lines indicate the boundaries of the stoichiometric flat window (La-to-Al flux ratio $\delta =$ 0.9 and 1.1).

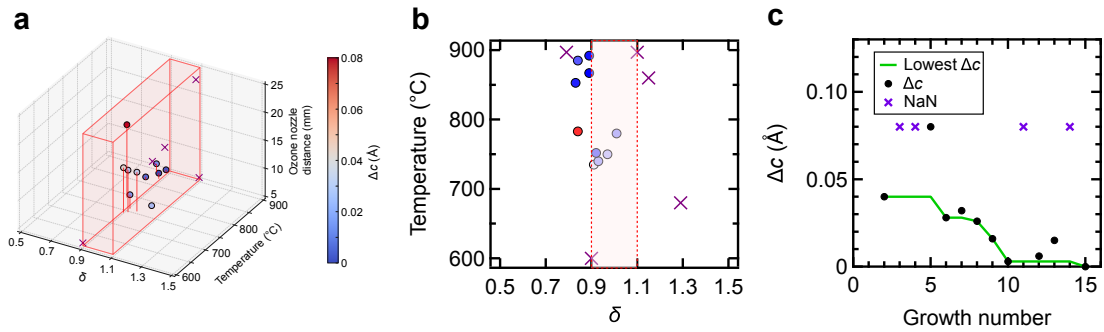

**Fig. 4. Optimization results. a** Experimental $\Delta c$ values in the three-dimensional parameter space for 15 observations. **b** The corresponding projection onto the La-to-Al flux ratio $\delta$-temperature plane. **c** Experimental $\Delta c$ values and the lowest experimental $\Delta c$ plotted as a function of growth number. In **a-c**, the purple crosses indicate the NaN (not a number) points at which the LAO phase was not obtained. In **a** and **b**, the red shaded regions indicate the stoichiometric flat window of $0.9 \leq \delta \leq 1.1$.

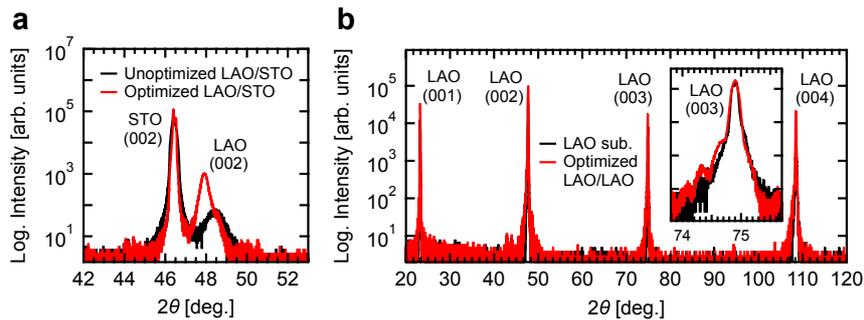

**Fig. 5. XRD characterizations. a** XRD $\theta$-$2\theta$ scans of the optimized and unoptimized LAO films grown on STO (001) substrates. **b** XRD $\theta$-$2\theta$ scans of the optimized LAO film grown on an LAO (001) substrate. In **b**, the inset shows the magnified image at the LAO (003) peak.

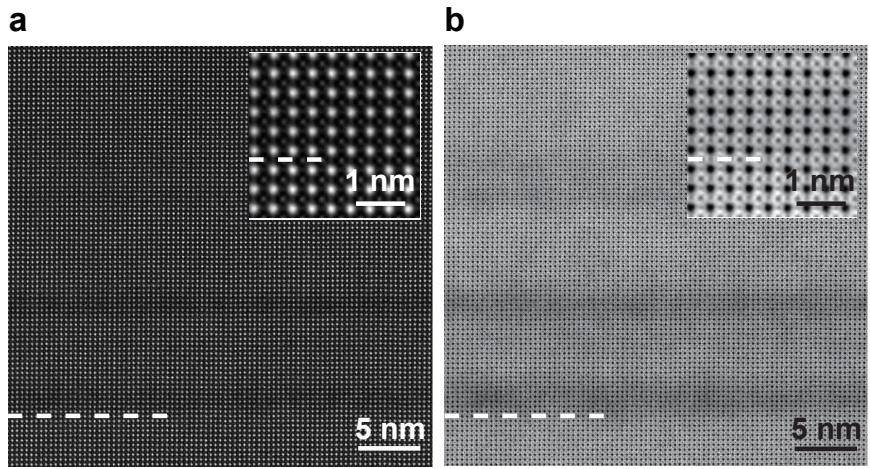

**Fig. 6. STEM characterizations.** (a) HAADF-STEM and (b) ABF-STEM images of the optimized LAO film grown on an LAO (001) substrate along the [100] direction. Dashed lines indicate the interfaces between the grown LAO layers and the substrates. Insets show magnified images at the interfaces.

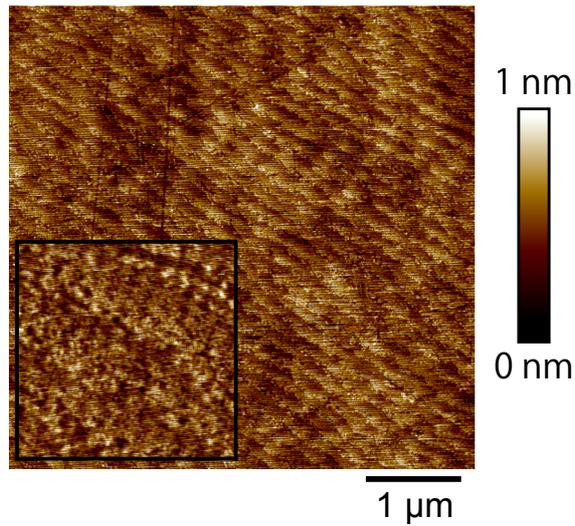

**Fig. 7. AFM characterizations.** AFM image of the 2-nm-thick optimized LAO film grown on an STO (001) substrate. The inset shows a magnified image highlighting the step and terrace structure.